# Room-Temperature High On/Off Ratio in Suspended Graphene Nanoribbon Field Effect Transistors

Ming-Wei Lin[1,*], Cheng Ling[1,*], Yiyang Zhang[1,2], Hyeun Joong Yoon[2], Mark Ming-Cheng Cheng[2], Luis A. Agapito[3], Nicholas Kioussis[3], Noppi Widjaja[1], and Zhixian Zhou[1, a)]

[1]Department of Physics and Astronomy, Wayne State University,
Detroit, MI 48201

[2]Department of Electrical and computer engineering, Wayne State University,
 Detroit, MI 48202

[3]Department of Physics, California State University, Northridge, CA 91330

Abstract

We have fabricated suspended few layer (1-3 layers) graphene nanoribbon field effect transistors from unzipped multiwall carbon nanotubes. Electrical transport measurements show that current-annealing effectively removes the impurities on the suspended graphene nanoribbons, uncovering the intrinsic ambipolar transfer characteristic of graphene. Further increasing the annealing current creates a narrow constriction in the ribbon, leading to the formation of a large band-gap and subsequent high on/off ratio (which can exceed $10^4$). Such fabricated devices are thermally and mechanically stable: repeated thermal cycling has little effect on their electrical properties. This work shows for the first time that ambipolar field effect characteristics and high on/off ratios at room temperature can be achieved in relatively wide graphene nanoribbon (15 nm ~50 nm) by controlled current annealing.

*These authors contributed equally.

a) Author to whom correspondence should be addressed, electronic mail: zxzhou@wayne.edu

**Introduction**

Graphene is a single atomic layer of graphite that exhibits exceptionally high carrier-mobility, offering the tantalizing possibility of graphene-based electronics [1]. In the mean time, graphene is a zero gap semiconductor with finite minimum conductivity, which poses a major problem for mainstream logic applications. One way to circumvent this problem is to slice graphene into nanometer-scale ribbons, where a band gap can be created by spatial confinement and edge-effects [2]. Electron-beam lithography was first used to pattern graphene nanoribbons (GNR) down to width <~20 nm for field-effect transistor (FET) applications; and a width-dependent transport gap was observed in these devices [3-5]. However, GNRs fabricated by electron beam lithography and subsequent oxygen plasma etching have relatively rough edges (on the order of few nanometers) limited by the resolution of electron beam lithography, which may degrade their electrical properties. Subsequently, several alternative methods have been developed to produce GNRs, including chemical sonication of exfoliated expandable graphite or chemically derived graphene sheets[6, 7], controlled nano-cutting with either metal particles or scanning probe tips[8-10], etching with physical masks (e.g. nanowires)[11], and longitudinal unzipping of multiwall carbon nanotubes [12-14]. Particularly, sub-10 nm GNRs with ultrasmooth edges have been produced by sonicating thermally exfoliated expandable graphite in solution [6, 15]. FET devices based on these GNRs have demonstrated an on/off ratio as high as $10^7$ at room temperature, representing a significant breakthrough in the field of graphene-based electronics [6] [15]. High on/off ratio has also been achieved in dual-gate bilayer graphene FETs, where a bandgap is created by applying a perpendicular electric field [16].

In this article, we present an alternative method to fabricate GNR-FETs that exhibit the characteristic ambipolar behavior and on/off ratio exceeding $10^4$. We fabricated FET devices consisting of a GNR suspended ~150 nm above the underneath Si substrate (which is used as the back gate). Subsequently, we used controlled current annealing to create a narrow constriction in the suspended GNR to open a confinement gap, thus to afford a high on/off ratio at room temperature. Atomic force microscopy (AFM) was used to confirm the formation of a narrow constriction in the GNRs. Room temperature high on/off ratio graphene transistors have also been previously realized by creating nanometer size quantum dots (QD) using electron-beam lithography and plasma over-etching [17]. However, these lithographically defined graphene QDs have an on-state conductance at least two orders of magnitude lower than that of our devices, and do not exhibit ambipolar behavior, presumably due to the substantial disorder induced by the adsorbed impurities and/or edge roughness [17].

**Experimental details**

The GNRs were produced by sonicating mildly-oxidized multiwall carbon nanotubes (MWNT) in a 1,2-dichloroethane (DCE) solution of poly(m-phenylenevinylene-co 2, 5-diy octocy- p-phenylenevinglene) (PmPV), where the PmPV is used as a surfactant to stabilize the unzipped GNRs in solution [14]. The solution was then centrifuged at 15000 rpm (Fisher Scientific Marathon 26kmr centrifuge) for 1 hr to remove aggregates and some of the remaining MWNTs; and a supernatant containing nanoribbons and remaining MWCTs was obtained. Next, the GNR samples from the supernatant were deposited on degenerately doped Si subtracts with 290 nm of thermal oxide, and subsequently non-contact mode AFM (Park System XE-70) was used to locate individual GNRs with respect to the prefabricated Au alignment marks and to characterize their thickness, width and length. The GNRs produced from this method mostly consist of 1-3 layers [14]. The AFM tip dilation effect (leading to artificial width increase) is accounted for based on the estimated tip radius provided by the tip manufacturer [18]. Standard electron-beam lithography (EBL) was used to pattern electrodes on selected GNRs followed by thermal metal deposition (a 0.5 nm Cr adhesion layer and 50 nm of Au) and lift-off in acetone. Suspension of the GNRs in FET devices was achieved by placing a small drop of 1:6 buffered hydrofluoric acid (HF) on top of the GNR device for 90 s to etch way approximately 150 nm of the $SiO_2$ underneath the ribbons [19, 20]. After wet etching, the device was transferred to hot isopropyl alcohol (which has low surface tension) and led try on a 120 °C hot plate. Finally, the device was annealed in vacuum at 600 °C to clean the ribbon surfaces and improve electrical contacts. In some of the studied devices, an additional EBL step was used to open a window in the electron-beam resist serving as etching mask for selectively etching the $SiO_2$ in the active device area. Both methods yielded similar results, with or without an additional EBL step. Surprisingly, about 50% of the over 30 devices with the suspended portion of the ribbon shorter than 800 nm survived the rather harsh fabrication processes.

The electrical transport properties of the suspended GNR devices were measured in high vacuum ($10^{-6}$ torr in a Lakeshore Cryogenics vacuum probe station) and at room temperature unless stated otherwise. A semiconductor parameter analyzer (Keithley 4200) was used to apply the annealing current and to measure the device characteristics. The degenerately doped Si substrate was used as a back gate. We repeatedly applied gradually increasing annealing current and subsequently carry out the electrical measurements *in situ* after every consecutive step. To avoid the possible collapse of the suspended GNRs, the back-gate voltage was limited to $-20 < V_g < +20$ V during the electrical measurements.

**Results and discussion**

We have studied over a dozen suspended few layer GNRs (1-2 nm thick) with a width between 15 and 50 nm. The length of the suspended GNR ranges from 100 nm to 800 nm. Figure 1(a) and (b) show atomic force microscopy (AFM) images of typical devices before and after etching, respectively. Line profile of the suspended GNR (Fig. 1c.) indicates that the ribbon is suspended ~150 nm above the surface of the remaining $SiO_2$ without substantial slacking (sagging).

Most of the suspended GNR devices used in this study showed ambipolar transfer characteristics with a charge neutrality point within a few volts from $V_g = 0$ after sufficient current annealing in vacuum (at a current density ~ $10^8$ A/cm$^2$). No obvious layer number dependence was observed in the transfer characteristics of our few layer (1-3 layers) GNRs likely due to the none AB (Bernal) stacking in these GNRs [21]. Upon further increasing the annealing current, about 30% of the suspended GNRs showed dramatic increase of the on/off ratio in conductance (or drain-source current) as defined by the value measured at $V_g = -15$ divided by the value at the charge neutrality point, while the rest of the GNRs were destroyed during the annealing processes likely due to localized overheating or electromigration [22]. Fig. 2 a shows the room-temperature conductance (G) versus gate voltage ($V_g$) for a representative suspended GNR device after sweeping the annealing bias-voltage from 0 to a predefined set-point of 2.9 V, and then decreasing the voltage back to 0 V, demonstrating the characteristic ambipolar behavior arising from the electron-hole symmetry of graphene. Fig. 2 b shows that the current versus bias-voltage (I-V) of the device is linear at low bias voltages, indicating near Ohmic electrical contacts. Further current annealing (by slightly raising the predefined voltage set-point) dramatically reduces the minimum conductance, while the on-state conductance is essentially unchanged (within a factor of two) as shown in Fig. 2 c. After the device is current annealed to 3.05 V, the on/off conductance ratio measured at $V_{ds} = 10$ mV approaches $10^6$. Fig. 2 d shows the current versus gate voltage (I vs. $V_g$) of the same device measured at different bias voltages after current annealing to 3.05 V, revealing an on/off ratio $> 10^4$ for 10 mV $\leq V_{ds} \leq$ 200 mV and $10^3$ ~$10^4$ for $V_{ds} = 500$ mV. Such high on/off ratios at room temperature have been previously observed in sub-10 nm GNRs, which were attributed to the opening of an effective band-gap at the order of hundreds meV primarily due to the confinement effect [6]. Since the band-gap decreases as the ribbon width increases and our ribbon is approximately 20 nm wide, the confinement-induced band-gap in our ribbon is expected to be of the order of ~10 meV [4, 23, 24].

To elucidate the origin of the large on/off ratio, we measured the G versus $V_g$ curve of the suspended GNR device in Fig. 2 at various temperatures (T) after it was annealed to 3.05 V. Fig. 3 a shows a semi-logarithmic plot of the minimum current ($I_{min}$) at the charge neutrality point versus 1/T, of the device. The data fit well to the thermal activation law: $G_{min} \sim \exp(-E_g/2k_BT)$ (where $k_B$ is the Boltzmann's constant), yielding a band-gap of $E_g \approx 0.6$ eV. A likely cause for such a large value of the band-gap and the consequent high on/off ratio is that further annealing beyond 2.9 V may have caused structural changes in the ribbon, such as creating a sub-10 nm constriction in the ribbon. In our suspended GNR devices, Joule annealing removes the impurities on the ribbon surfaces [25], leading to the ambipolar behavior with the charge neutrality point occurring at $Vg \approx 0$ V (Fig. 2 a). Further annealing (increasing the pre-defined voltage set-point) facilitates structural reconstruction in the suspended ribbon, especially near the edges [26]. To confirm that annealing beyond 2.9 V created a narrow constriction in the ribbon, it is necessary to characterize the structure of the suspended GNRs especially at the end of the current annealing. Raman spectroscopy has proven to be a powerful and noninvasive tool to characterize the structures of graphene [27-30], however it lacks the spatial resolution that is needed to confirm the formation of a nanoscale constriction. While high resolution transmission electron microscopy ( TEM ) is capable of visualizing individual carbon atoms [31], it is difficult to directly characterize the structure of our GNR devices using TEM as they are fabricated on $Si/SiO_2$ substrates. In this study, we used AFM to characterize the suspended GNR device along with another suspended GNR device fabricated from an adjacent section of the same ribbon as shown in Fig. 3b. The two suspended sections are thus expected to have the same width, thickness and comparable length; except that no further current annealing was carried out in the lower section as soon as the characteristic ambipolar behavior with a low on/off ratio (< 10) was observed. Fig. 3 c and d show the high resolution AFM images of two sections of the same ribbon with high and low on/off ratios, respectively, clearly indicating that i) both sections of the GNR are suspended without sagging (bowing); and ii) there is a notch near the middle of the section of the ribbon with high on/off ratio (Fig. 3c) while the GNR section with low on/off ratio is highly uniform (Fig. 3d). The notch in the AFM image is a clear indication that current annealing beyond 2.9 V created a narrow constriction in the suspended GNR, resulting in the large on/off ratio. It is also worth pointing out that the AFM images in Fig. 3 were taken after electrical measurements with a gate voltage swept between -15 and 15V and several thermal cycles between a cryogenic temperature (4.3 K or 77 K) and room temperature, demonstrating that our suspended GNRs are thermally and mechanically stable (even that GNR with a narrow constriction).

Similar transfer characteristics and high on/off ratios have been observed in three other suspended GNR devices with varying width, length and thickness. Fig. 4a shows the transfer characteristics of a device consisting of a suspended GNR ~450 nm long, ~45 nm wide, and ~1.6 nm thick. We note that before current annealing the suspended ribbon was p-doped with a charge neutrality point beyond +15 V, which can be partially attributed to the adsorption of air or water molecules, or PMMA residue [32, 33]. After intermediate current annealing, the GNR exhibits ambipolar behavior with the minimum conductance associated with the charge neutrality point shifting to $V_g \approx 0$ V, indicating that the adsorbed charge impurities have been largely removed by current annealing. Further annealing decreases the minimum conductance by three orders of magnitude, while the overall transfer characteristics and the on-state conductance remain essentially unchanged. Fig. 4 b shows that the transfer characteristics of the device remains virtually the same after number of thermal cycles, although the unintentional doping level changes after each thermal cycle as indicated by the shift of the minimum conductance along the horizontal axis. When the conductance G is plotted versus ($V_g$-$V_{G-min}$) in Fig. 4c, all the three curves collapse into a single one, indicating again that the electrical properties of our devices are robust. To further demonstrate the good thermal and mechanical stability of our high on/off ration devices, we measured the electrical properties of a third device (consisting of a GNR ~ 19 nm wide and ~1.2 nm thick) before and after an additional *ex situ* annealing step at 600 $^O$C in a vacuum furnace. Fig. 4 d shows that the transfer characteristic of the device remains essentially the same after the additional *ex situ* annealing step.

High on/off ratios have been observed in GNRs of comparable width (tens of nanometers wide) but usually at cryogenic temperatures and have been attributed to the opening of a transport gap [4, 34-37]. Several possible mechanisms have been proposed to explain the large transport gap observed in GNR-FETs at low temperatures, ranging from re-normalized lateral confinement due to localized edge states [4, 34], to percolation driven metal-insulator-transition caused by charged impurities [35], to quasi-one-dimensional Anderson localization [38], and to Coulomb blockade due to edge-roughness [39]. More recent experimental studies on disordered GNRs further indicate that charge transport in the conduction gap of GNRs is likely dominated by localized states [36] and/or isolated charge puddles acting as quantum dots [37]. These mechanisms may partially contribute to the high on/off ratio in our devices. However, they are unlikely the primary cause, since the over $10^4$ on/off ratio in our devices was observed at room temperature and only after sufficient current annealing. For instance, when the device in Fig. 2 is current annealed up to 2.9 V, the on/off ratio is less than 2. The large on/off ratio in the device was obtained only after it was current annealed by applying a bias voltage beyond 2.9 V.

Based on the results of AFM characterization of the suspended GNRs (Fig. 3), we suggest that the formation of a constriction, at the critical annealing current, as the most likely origin of the high on/off ratio. Fig. 5 a-c schematically shows a relatively wide uniform GNR ($GNR_0$), the same GNR with a narrow constriction, which effectively forms a finite ribbon ($GNR_1$), and a narrow uniform GNR ($GNR_2$) with a descending width ($GNR_0 > GNR_1 > GNR_2$). Because of the stronger confinement in the narrow constriction ($GNR_1$) compared to the uniform wide GNR ($GNR_0$), the bandgap for $GNR_1$ (Fig. 5 e) is expected to be larger than that for $GNR_0$ (Fig. 5 d). In the same fashion, the bandgap of $GNR_2$ (Fig. 5 f) is larger than that of $GNR_0$ (Fig. 5 d) due to its narrower width [3-5]. In addition to the lateral confinement, region $GNR_1$ is also longitudinally confined by regions $QD_1$ and $QD_2$. The longitudinal confinement further increases the bandgap of the otherwise infinite ribbon $GNR_1$ to a value that can be possibly larger than that for the narrower ribbon $GNR_2$, thus resulting in higher on/off ratios. Therefore, the double-confinement picture seems to be responsible for the high on/off ratios which are comparable to those of possibly much narrower (sub-5nm) ribbons [40]. More quantitative understanding of the large bandgap requires detailed information on the dimensions and even the edge structures of the nano-constrictions, which is beyond the scope of this work. Recently, bandgaps of ~ 50 meV have been observed in relatively wide (~ 100 nm) annealed nanotube-derived GNRs [41], indicating that it is possible to obtain a bandgap ~ 500 meV in GNRs of ~ 10 nm wide. However, the bandgap of these GNRs was substantially larger than that of much narrower GNRs synthesized using a similar approach; and the difference in edge structures was suggested as the primary cause of the discrepancy [14, 41]. Unfortunately, the lack of edge information on these GNRs prevents the definitive identification of the true origin of the discrepancy.

Discrete conductance peaks, superimposed on the main G-vs-$V_g$ curve, were observed at low temperatures in some annealed samples (Fig. 6). These conductance oscillations can be attributed to the quantized energy levels of regions $QD_1$ and $QD_2$ in Fig. 5b, which because of lack of any periodicity or quasi-periodicity act as two quantum dots in series. Therefore, these conductance peaks are due to resonant tunneling through the quantized levels of the dots (shown in red in Fig. 5 e) tuned by the applied gate voltage. The spacing and periodicity of such peaks depend on the size and symmetry of the quantum dots. Asymmetry between the discrete energy levels of $QD_1$ and $QD_2$ leads to lack of resonance and random cancellation of the transmission through some levels [17].

During the course of this work, we became aware that room-temperature high on/off ratios were also observed in graphene nano-constrictions created by first forming a constriction in the gold etch mask covering the graphene channel (using electromigration ) and subsequent plasma etching of the underneath graphene [42]. However, the nano-constrictions presented in this work are created without a mask, which may lead to much cleaner graphene nano-constrictions with substantially lower disorder.

**Conclusion**

We have fabricated suspended GNR-FET devices from GNRs derived from high quality MWNTs. By controlled current-annealing of the suspended GNRs, a large band gap at the order of hundreds meV can be created in GNRs with varying width and thickness, leading to orders of magnitude increase of on/off ratio at room temperature. The formation of such a large band gap can be largely attributed to the creation of a narrow constriction in the suspended GNR as confirmed by AFM. Furthermore, the suspended GNRs and their narrow constrictions are structurally robust, and the electrical properties of the devices remain unchanged after number of thermal cycles. The mask-free fabrication method presented in this study also creates opportunities for studying ultraclean graphene quantum dots.

Acknowledgement

Z.Z acknowledges the support of the Wayne State University startup funds. Part of this research was conducted at the Center for nanophase Materials Sciences under project # CNMS2009-044. The authors also thank Boris Nadgorny and Raghu Murali for helpful discussions.

[1] Novoselov K. S., Geim A. K., Morozov S. V., Jiang D., Katsnelson M. I., Grigorieva I. V., Dubonos S. V. and Firsov A. A. 2005 Two-dimensional gas of massless dirac fermions in graphene *Nature* **438** 197

[2] Son Y.-W., Cohen M. L. and Louie S. G. 2006 Energy gaps in graphene nanoribbons *Phys. Rev. Lett.* **97** 216803

[3] Han M. Y., Ozyilmaz B., Zhang Y. and Kim P. 2007 Energy band-gap engineering of graphene nanoribbons *Phys. Rev. Lett.* **98** 206805

[4] Chen Z. H., Lin Y. M., Rooks M. J. and Avouris P. 2007 Graphene nano-ribbon electronics *Physica E* **40** 228

[5] Berger C., Song Z., Li X., Wu X., Brown N., Naud C., Mayou D., Li T., Hass J., Marchenkov A. N., Conrad E. H., First P. N. and de Heer W. A. 2006 Electronic confinement and coherence in patterned epitaxial graphene *Science* **312** 1191

[6] Li X., Wang X., Zhang L., Lee S. and Dai H. 2008 Chemically derived, ultrasmooth graphene nanoribbon semiconductors *Science* **319** 1229

[7] Wu Z. S., Ren W., Gao L., Liu B., Zhao J. and Cheng H.-M. 2010 Efficient synthesis of graphene nanoribbons sonochemically cut from graphene sheets *Nano Res.* **3** 16

[8] Datta S. S., Strachan D. R., Khamis S. M. and Johnson A. T. C. 2008 Crystallographic etching of few-layer graphene *Nano Lett.* **8** 1912

[9] Ci L., Xu Z., Wang L., Gao W., Ding F., Kelly K., Yakobson B. and Ajayan P. 2008 Controlled nanocutting of graphene *Nano Research* **1** 116

[10] Tapaszto L., Dobrik G., Lambin P. and Biro L. P. 2008 Tailoring the atomic structure of graphene nanoribbons by scanning tunnelling microscope lithography *Nat Nano* **3** 397

[11] Bai J., Duan X. and Huang Y. 2009 Rational fabrication of graphene nanoribbons using a nanowire etch mask *Nano Lett.* **9** 2083

[12] Kosynkin D. V., Higginbotham A. L., Sinitskii A., Lomeda J. R., Dimiev A., Price B. K. and Tour J. M. 2009 Longitudinal unzipping of carbon nanotubes to form graphene nanoribbons *Nature* **458** 872

[13] Jiao L., Zhang L., Wang X., Diankov G. and Dai H. 2009 Narrow graphene nanoribbons from carbon nanotubes *Nature* **458** 877

[14] Jiao L., Wang X., Diankov G., Wang H. and Dai H. 2010 Facile synthesis of high-quality graphene nanoribbons *Nat Nano* **5** 321

[15] Wang X., Ouyang Y., Li X., Wang H., Guo J. and Dai H. 2008 Room-temperature all-semiconducting sub-10-nm graphene nanoribbon field-effect transistors *Phys. Rev. Lett.* **100** 206803

[16] Xia F., Farmer D. B., Lin Y.-m. and Avouris P. Graphene field-effect transistors with high on/off current ratio and large transport band gap at room temperature *Nano Letters* **10** 715

[17] Ponomarenko L. A., Schedin F., Katsnelson M. I., Yang R., Hill E. W., Novoselov K. S. and Geim A. K. 2008 Chaotic dirac billiard in graphene quantum dots *Science* **320** 356

[18] Miriam Moreno-Moreno A. C.-G., Gabino Rubio-Bollinger, Julio Gomez-Herrero, Nicolás Agraït, 2009 Ultralong natural graphene nanoribbons and their electrical conductivity *Small* **5** 924

[19] Bolotin K. I., Sikes K. J., Jiang Z., Klima M., Fudenberg G., Hone J., Kim P. and Stormer H. L. 2008 Ultrahigh electron mobility in suspended graphene *Solid State Commun.* **146** 351

[20] Du X., Skachko I., Barker A. and Andrei E. Y. 2008 Approaching ballistic transport in suspended graphene *Nat Nano* **3** 491

[21] Sagar A., Lee E. J. H., Balasubramanian K., Burghard M. and Kern K. 2009 Effect of stacking order on the electric-field induced carrier modulation in graphene bilayers *Nano Letters* **9** 3124

[22] Murali R., Yang Y., Brenner K., Beck T. and Meindl J. D. 2009 Breakdown current density of graphene nanoribbons *Applied Physics Letters* **94** 243114

[23] Lin Y.-M., Perebeinos V., Chen Z. and Avouris P. 2008 Electrical observation of subband formation in graphene nanoribbons *Phys. Rev. B* **78** 161409

[24] Lian C., Tahy K., Fang T., Li G., Xing H. G. and Jena D. 2010 Quantum transport in graphene nanoribbons patterned by metal masks *Appl. Phys. Lett.* **96** 103109

[25] Bolotin K. I., Sikes K. J., Hone J., Stormer H. L. and Kim P. 2008 Temperature-dependent transport in suspended graphene *Phys. Rev. Lett.* **101** 096802

[26] Jia X., Hofmann M., Meunier V., Sumpter B. G., Campos-Delgado J., Romo-Herrera J. M., Son H., Hsieh Y.-P., Reina A., Kong J., Terrones M. and Dresselhaus M. S. 2009 Controlled formation of sharp zigzag and armchair edges in graphitic nanoribbons *Science* **323** 1701

[27] Cong C., Yu T. and Wang H. 2010 Raman study on the g mode of graphene for determination of edge orientation *ACS Nano* **4** 3175

[28] Ren W., Saito R., Gao L., Zheng F., Wu Z., Liu B., Furukawa M., Zhao J., Chen Z. and Cheng H.-M. 2010 Edge phonon state of mono- and few-layer graphene nanoribbons observed by surface and interference co-enhanced raman spectroscopy *Phys. Rev. B* **81** 035412

[29] Casiraghi C., Hartschuh A., Qian H., Piscanec S., Georgi C., Fasoli A., Novoselov K. S., Basko D. M. and Ferrari A. C. 2009 Raman spectroscopy of graphene edges *Nano Lett.* **9** 1433

[30] Begliarbekov M., Sul O., Kalliakos S., Yang E.-H. and Strauf S. 2010 Determination of edge purity in bilayer graphene using mu-raman spectroscopy *Appl. Phys. Lett.* **97** 031908

[31] Girit C. O., Meyer J. C., Erni R., Rossell M. D., Kisielowski C., Yang L., Park C.-H., Crommie M. F., Cohen M. L., Louie S. G. and Zettl A. 2009 Graphene at the edge: Stability and dynamics *Science* **323** 1705

[32] Moser J., Verdaguer A., Jimenez D., Barreiro A. and Bachtold A. 2008 The environment of graphene probed by electrostatic force microscopy *Appl. Phys. Lett.* **92** 123507

[33] Dan Y., Lu Y., Kybert N. J., Luo Z. and Johnson A. T. C. 2009 Intrinsic response of graphene vapor sensors *Nano Letters* **9** 1472

[34] Han M. Y., Ozyilmaz B., Zhang Y. B. and Kim P. 2007 Energy band-gap engineering of graphene nanoribbons *Phys. Rev. Lett.* **98** 206805

[35] Adam S., Cho S., Fuhrer M. S. and Sarma S. D. 2008 Density inhomogeneity driven percolation metal-insulator transition and dimensional crossover in graphene nanoribbons *Phys. Rev. Lett.* **101** 046404

[36] Han M. Y., Brant J. C. and Kim P. 2010 Electron transport in disordered graphene nanoribbons *Phys. Rev. Lett.* **104** 056801

[37] Gallagher P., Todd K. and Goldhaber-Gordon D. 2010 Disorder-induced gap behavior in graphene nanoribbons *Phys. Rev. B* **81** 115409

[38] Evaldsson M., Zozoulenko I. V., Hengyi X. and Heinzel T. 2008 Edge-disorder-induced anderson localization and conduction gap in graphene nanoribbons *Phys. Rev. B* **78** 161407

[39] Sols F., Guinea F. and Neto A. H. C. 2007 Coulomb blockade in graphene nanoribbons *Phys. Rev. Lett.* **99** 166803

[40] Xie L., Jiao L. and Dai H. Selective etching of graphene edges by hydrogen plasma *Journal of the American Chemical Society* **132** 14751

[41]  ShimizuT, HaruyamaJ, Marcano D. C., Kosinkin D. V., Tour J. M., HiroseK and SuenagaK Large intrinsic energy bandgaps in annealed nanotube-derived graphene nanoribbons *Nat Nano* **6** 45

[42]  Lu Y., Goldsmith B., Strachan D. R., Lim J. H., Luo Z. and Johnson A. T. C. High-on/off-ratio graphene nanoconstriction field-effect transistor *Small* **6** 2748

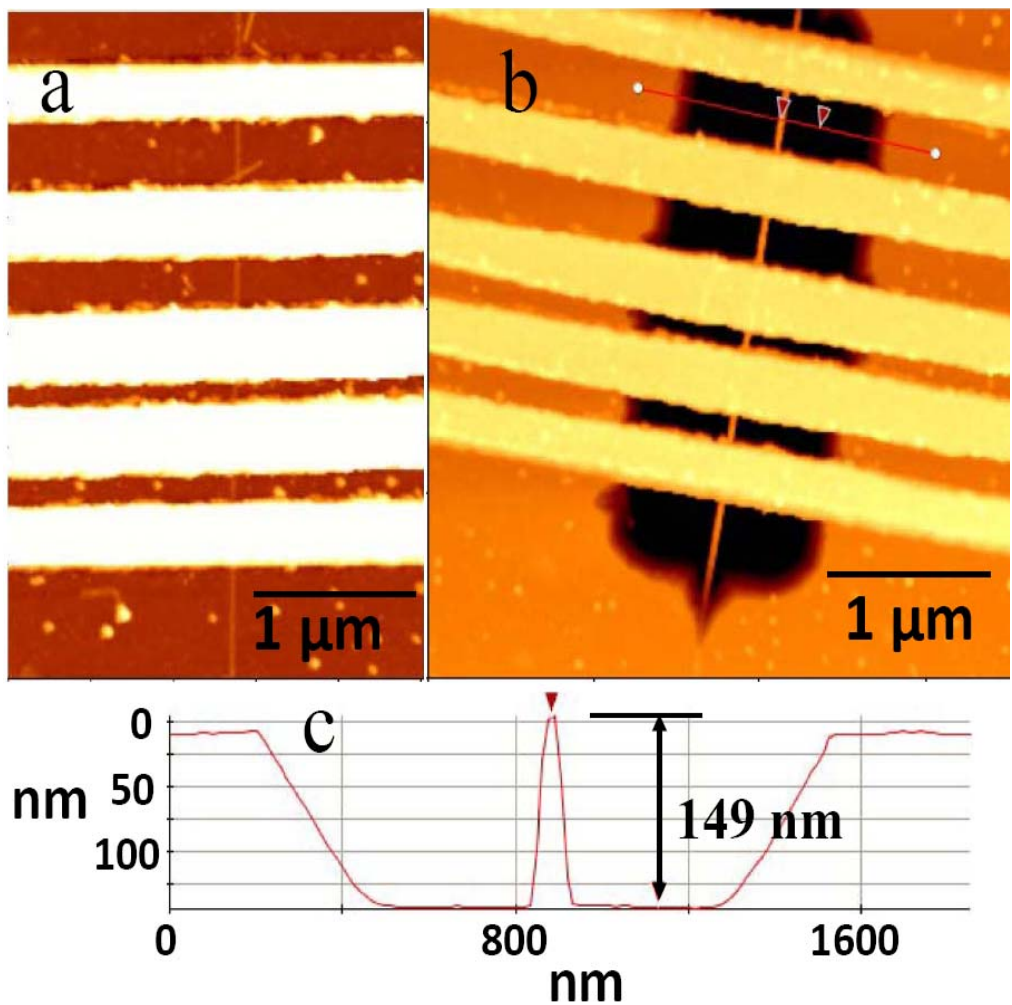

Fig.1. AFM images of typical FET devices consisting of a GNR contacted by Au electrodes before (a) and after (b) suspending the GNR. (c) Line profile of the top section of the suspended GNR.

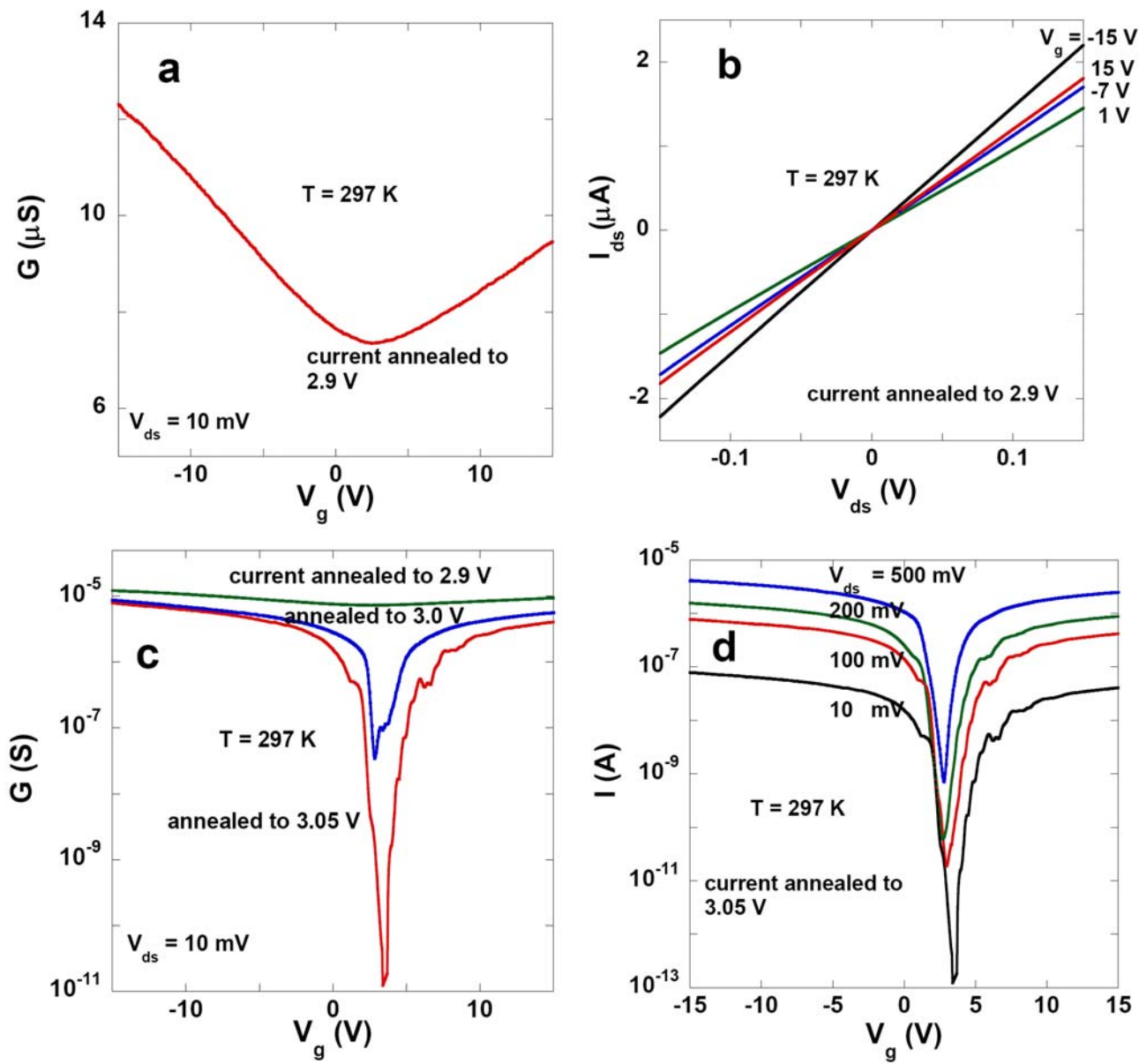

Fig. 2. Electrical transport properties of a representative suspended GNR FET device measured at room temperature, where the suspended GNR is 21±3 nm wide, ~1.4 nm thick and ~600 nm long. (a) Transfer characteristic of the device after current annealing to a predefined bias voltage set-point of 2.9 V. (b) I-V characteristics of the device measured at various gate voltages ranging from -15 to 15 V. (c) Transfer characteristic measured after various degrees of current annealing. (d) Current versus gate voltage ($V_g$) measured at different bias voltages after the final stage of current annealing (annealed to 3.05 V).

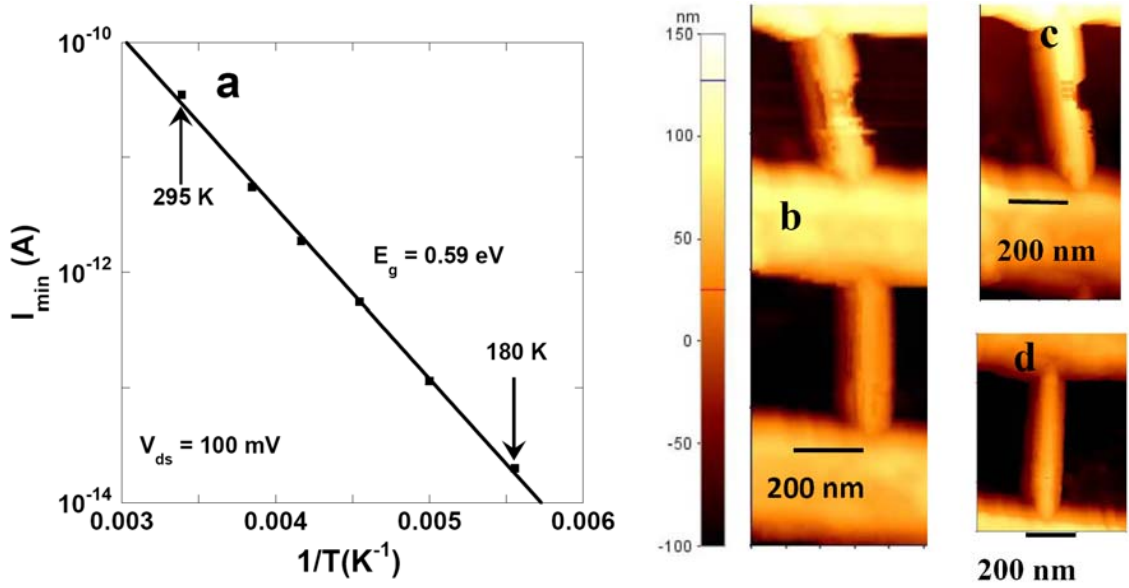

Fig. 3. (a) Temperature dependence of the minimum current at the charge neutrality point measured at $V_{ds}$ = 100 mV for the suspended GNR device in Figure 2. (b) Top part: AFM image of the same GNR in (a); bottom part: another section of the suspended GNR. (c-d) Higher resolution AFM images of the top and bottom sections of the GNR in (b), respectively.

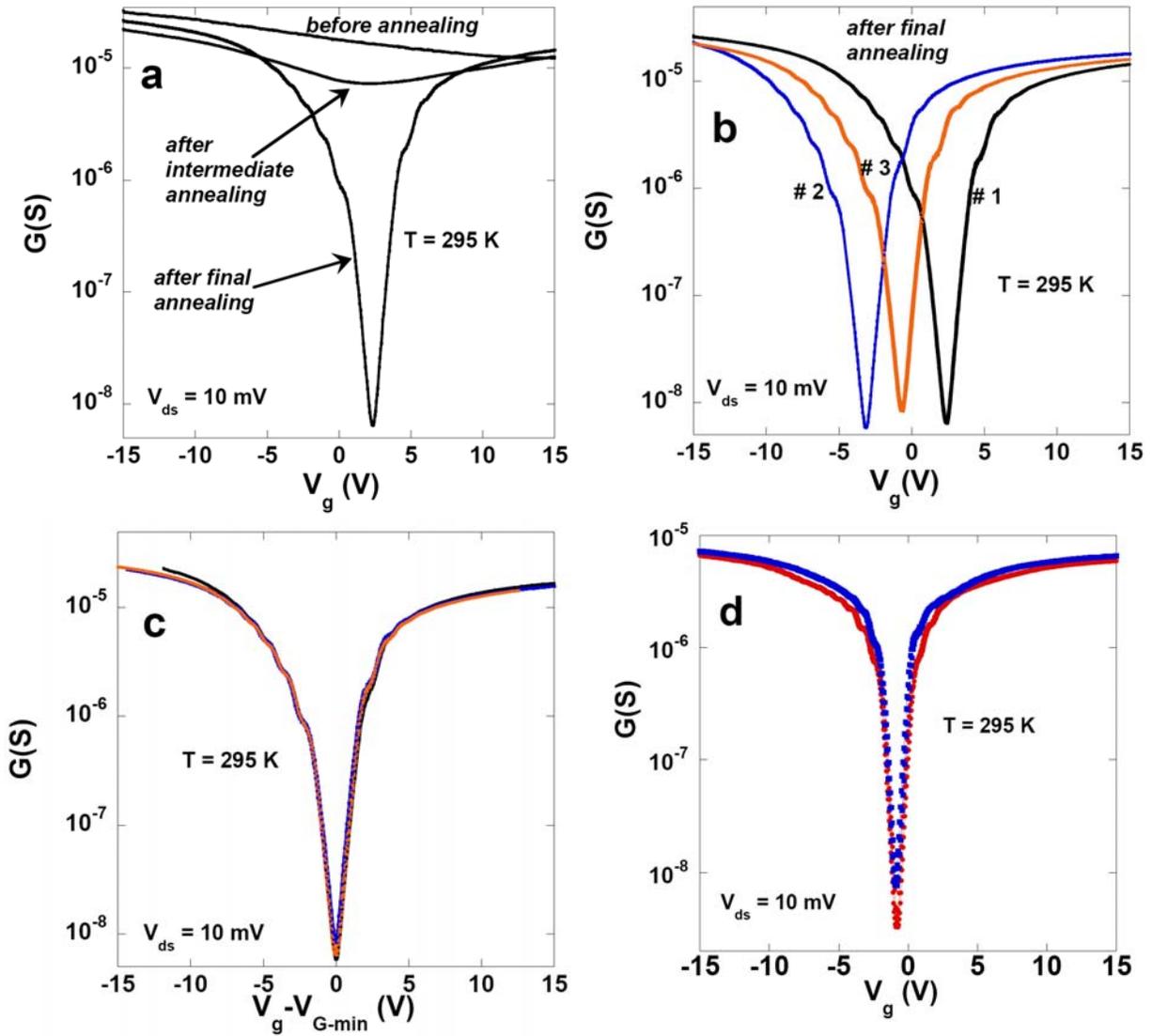

Fig. 4. (a-c) Transfer characteristics of a second suspended GNR FET device measured at room temperature; the suspended GNR is 45±3 nm wide, ~1.6 nm thick and ~450 nm long. (a) Transfer characteristic measured after various degrees of annealing. (b) Conductance versus gate voltage measured at room temperature after the final current annealing stage and after different number of thermal cycles; # 1, #2, and #3 next to the corresponding $G(V_g)$ curves represent measurements after 1, 2 and 3 thermal cycles, respectively. (c) Gate dependent conductance data in (b) plotted as a function of $V_g-V_{G-min}$. (d) Room temperature transfer characteristics of a third suspended GNR device (19±3 nm wide, ~1.2 nm thick, and ~300 nm long) measured before (blue solid squires) and after (red solid dots) *ex situ* annealing at 600 °C for 10 minutes in a vacuum furnace.

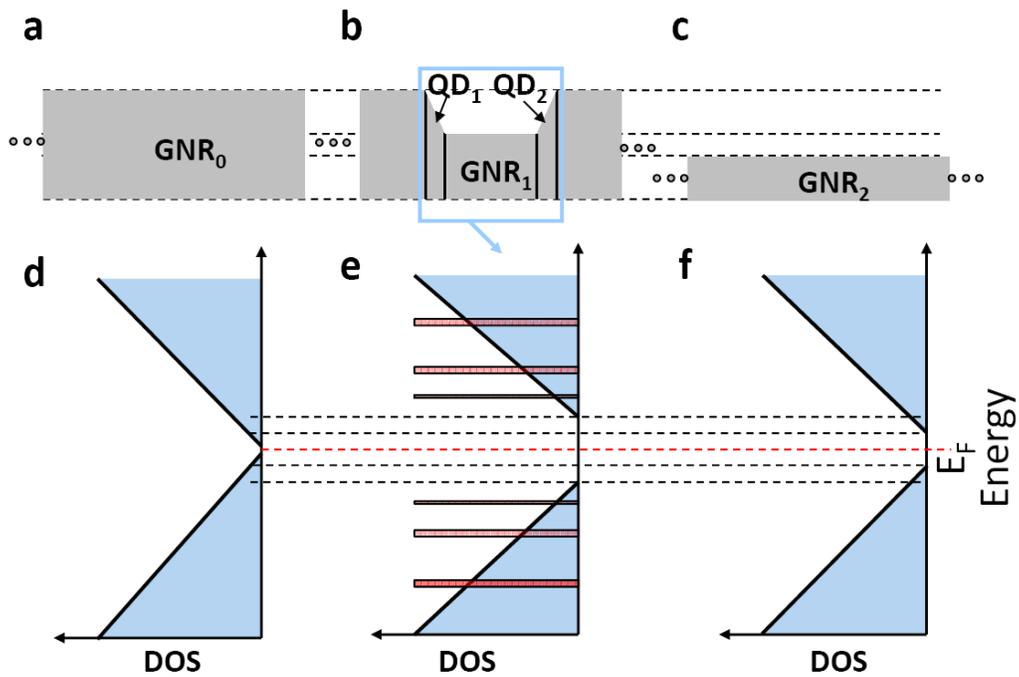

Fig. 5. Schematic density of states (DOS) for GNRs. (a) and (c) represent two infinite GNRs of different widths; the corresponding energy bandgaps, observed in the DOS panels (d) and (f), follow the known inverse relation with ribbon width. (b) GNR with a constriction. (e) DOS projected on the region of the constriction; the quantized energy levels of quantum dots $QD_1$ and $QD_2$ are shown in red and the states of region $GNR_1$ are shown in blue.

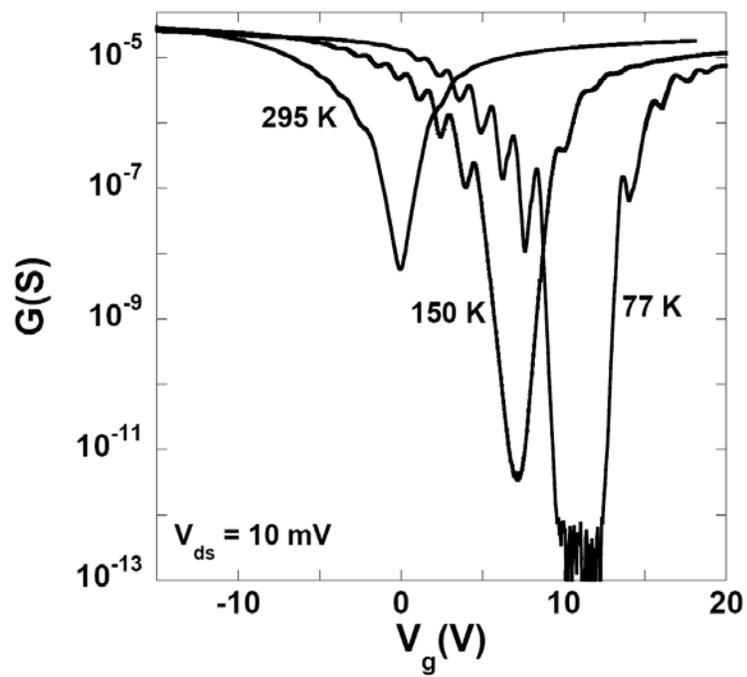

Fig. 6 Conductance versus gate voltage of the GNR device in Fig. 4d measured at 77, 150, and 295 K.